\pdfoutput=1
\documentclass[12pt,tikz]{article}

\usepackage{amsmath}
\usepackage{amssymb}
\usepackage{graphicx}
\usepackage{xcolor}
\numberwithin{equation}{section}
\usepackage{array}
\usepackage{mathtools}
\usepackage{dsfont}
\usepackage{mathrsfs}
\usepackage{tikz}\usetikzlibrary{matrix,fit}
\usepackage{varwidth}
\usepackage{enumerate}

\usepackage[margin=1in]{geometry}

\usepackage{mathabx}
\usepackage{empheq}

\usepackage{float}

\usepackage{hyperref}
\usepackage{setspace}

\usepackage{slashed}
\usepackage{upgreek}
\usepackage{appendix}

\usepackage{tikz-cd}
\usetikzlibrary{positioning,backgrounds,patterns}
\usetikzlibrary{arrows,calc,through,backgrounds,matrix,decorations.pathmorphing}
\usepackage{pst-node}
\usepackage{pstricks, pst-poly}
\usepackage{pgfkeys}
\usetikzlibrary{patterns.meta}
\usepgflibrary{patterns}
\usetikzlibrary{patterns}

\usepackage{tabstackengine}
\fixTABwidth{T}

\newdimen\mytextwidth
\newcommand\rem[2][cyan!40!green]{\noindent\nobreak\hfil\penalty1000\hfilneg
\mytextwidth=\linewidth\advance\mytextwidth by 2mm%
\begin{tikzpicture}[baseline=-\the\dimexpr\fontdimen22\textfont2\relax]\node[outer sep=0pt,draw=black,fill=#1,fill opacity=1,text opacity=1,rectangle,rounded corners]{\begin{varwidth}{\mytextwidth}\textcolor{white}{#2}\end{varwidth}};
\end{tikzpicture}\allowbreak%
}

\newcommand\whiterem[2][white!]{\noindent\nobreak\hfil\penalty1000\hfilneg
\mytextwidth=\linewidth\advance\mytextwidth by 2mm%
\begin{tikzpicture}[baseline=-\the\dimexpr\fontdimen22\textfont2\relax]\node[outer sep=0pt,draw=black,fill=#1,fill opacity=1,text opacity=1,rectangle,rounded corners,line width=1.5pt]{\begin{varwidth}{\mytextwidth}\textcolor{black}{#2}\end{varwidth}};
\end{tikzpicture}\allowbreak%
}

\newcommand{\dd}{\partial}

\newcommand{\CP}{\mathbb{CP}}
\newcommand{\CC}{\mathbb{C}}

\newcommand{\bea}{\begin{equation}}
\newcommand{\eea}{\end{equation}}
\newcommand{\bear}{\begin{eqnarray}}
\newcommand{\eear}{\end{eqnarray}}
\newcommand{\bearr}{\begin{eqnarray*}}
\newcommand{\eearr}{\end{eqnarray*}}

\newcommand{\appendixnumberline}[1]{Appendix #1.\space}

\let\oldappendix\appendix
\makeatletter
\renewcommand{\appendix}{%
  \addtocontents{toc}{\let\protect\numberline\protect\appendixnumberline}%
  \renewcommand{\@seccntformat}[1]{\large\bfseries Appendix . }%
  \oldappendix
}
\makeatother

\usepackage{mdframed}

\newmdenv[
  topline=false,
  bottomline=false,
  rightline=false,
  linewidth=2pt,
  skipabove=\topsep,
  skipbelow=\topsep
]{siderules}

\newmdenv[
  topline=false,
  bottomline=false,
  linewidth=2pt,
  skipabove=\topsep,
  skipbelow=\topsep
]{siderulesright}

\makeatletter
\renewcommand{\@seccntformat}[1]{\csname the#1\endcsname.\quad}
\makeatother

\usepackage{setspace}

\usepackage{xpatch}

\makeatletter
\renewcommand{\@chap@pppage}{%
  \clear@ppage
  \thispagestyle{plain}%
  \if@twocolumn\onecolumn\@tempswatrue\else\@tempswafalse\fi
  \null\vfil
  \markboth{}{}%
  {\centering
   \interlinepenalty \@M
   \normalfont
   \MakeUppercase \appendixpagename\par}%
  \if@dotoc@pp
    \addappheadtotoc
  \fi
  \vfil\newpage
  \if@twoside
    \if@openright
      \null
      \thispagestyle{empty}%
      \newpage
    \fi
  \fi
  \if@tempswa
    \twocolumn
  \fi
}
\makeatother

\usepackage{tocloft}

\let \savenumberline \numberline
\def \numberline#1{\savenumberline{#1.}}

\usepackage{etoolbox}
\patchcmd{\tableofcontents}{\@starttoc}{\vspace{-0.3cm}\@starttoc}{}{}

\usepackage{hyperref}

\usepackage{titlesec}

\titleformat*{\section}{\large\bfseries}
\titleformat*{\subsection}{\normalsize\bfseries}
\titleformat*{\subsubsection}{\normalsize\bfseries}
\titleformat*{\paragraph}{\large\bfseries}
\titleformat*{\subparagraph}{\large\bfseries}
\titlespacing{\author}{-5pt}{-5pt}{-5pt}[-5pt]

\makeatletter
\renewcommand\subsubsection{\@startsection{subsubsection}{3}{\z@}%
                                     {-3.25ex\@plus -1ex \@minus -.2ex}%
                                     {-1.5ex \@plus -.2ex}
                                     {\normalfont\normalsize\bfseries}}
\renewcommand\subsection{\@startsection{subsection}{3}{\z@}%
                                     {-3.25ex\@plus -1ex \@minus -.2ex}%
                                     {-1.5ex \@plus -.2ex}
                                     {\normalfont\normalsize\bfseries}}                                     
\makeatother

\usepackage{slashed}

\usepackage{accents}
\newcommand\thickbar[1]{\accentset{\rule{.7em}{.8pt}}{#1}}

   \usepackage{stmaryrd}
   
   \definecolor{grey}{rgb}{0.5,0.5,0.5}
   
   \DeclareFontFamily{U}{solomos}{}
\DeclareErrorFont{U}{solomos}{m}{n}{10}
\DeclareFontShape{U}{solomos}{m}{n}{
  <-> s*[1.1]  gsolomos8r
}{}
   
\newcommand{\vkappa}{\text{\usefont{U}{solomos}{m}{n}\symbol{'153}}}   

\begin{document}

\title{\vspace{-1.0cm} Sigma models as Gross-Neveu models. II\footnote{Prepared for the A.A.~Slavnov memorial volume; the memorial conference took place on 21-22 December 2022 at Steklov Mathematical Institute.}}
\author{Dmitri Bykov\footnote{Emails:
 bykov@mi-ras.ru, dmitri.v.bykov@gmail.com}
\\  \vspace{-0.3cm}  \\
{ \small
 $\bullet$ Steklov Mathematical Institute of Russian Academy of Sciences,}\\
{\small Gubkina str. 8, 119991 Moscow, Russia}
\\
{\small $\bullet$ Institute for Theoretical and Mathematical Physics,}\\{\small 
Lomonosov Moscow State University, 119991 Moscow, Russia}}

\date{}

{\let\newpage\relax\maketitle}

\maketitle

\vspace{-0.5cm}
\textbf{Abstract.} We summarize some (mostly geometric) facts underlying the relation between 2D integrable sigma models and generalized Gross-Neveu models, emphasizing connections to the theory of nilpotent orbits, Springer resolutions and quiver varieties. This is meant to shed light on the general setup when this correspondence holds.

\begin{flushright}\emph{To the memory of A.~A.~Slavnov, with gratitude}
\end{flushright}

\vspace{0.5cm}
The goal of this paper is to summarize some ideas pertaining to the recently discovered relation between integrable sigma models in 2D and nilpotent (co)adjoint orbits of complex Lie groups (cf.~\cite{Bykov1, Bykov2} and references therein). Even if not immediately obvious, the latter feature in physics in the context of Gross-Neveu (GN) models. We mostly concentrate on the classical theory\footnote{Classical integrability of Gross-Neveu models was discovered in~\cite{Papanicolaou, ZM}; see also~\cite{DunneThies, Thies, Ashwinkumar} for more recent developments.}, apart from a brief discussion of quantization in the context of 1D models. As we shall see, the classical theory is rich in itself, exhibiting connections between sigma models, nilpotent orbits, resolutions of their closures and quiver formulations thereof. By now it is fairly clear that the relation is rather robust and may eventually lead to a better understanding of the quantum theory of sigma models, though some of the steps involved still remain conjectures. 

\section{Sigma models in mechanics}

We start with the 1D prototype of the models that we plan to study. Let $x: \mathbb{R}_t \mapsto (\mathcal{M}, \vkappa)$ be the trajectory of a particle moving on a Riemannian manifold $\mathcal{M}$ with metric~$\vkappa$. Free particle motion is equivalent to the problem of finding geodesics on $\mathcal{M}$, and is described by the following action:
\bea
\mathcal{S}=\int\,dt\,{1\over 2}\,\vkappa_{ij}\,\dot{x}^i \dot{x}^j
\eea
Alternatively, we may describe it in the Hamiltonian formalism by introducing a symplectic manifold $(T^\ast \mathcal{M}, \,\Omega=\sum_i\,dp_i \wedge dx^i)$, and the corresponding trajectory $y: \mathbb{R}_t \mapsto T^\ast\mathcal{M}$. In this case the action may be rewritten as
\bear
&&\mathcal{S}=\int\,dt\,\left(\,p_i \, \dot{x}^i-\mathsf{H}(x, p) \right)\\ \label{metrham}
&& \mathsf{H}(x, p)={1\over 2}\,\vkappa^{ij}\,p_i p_j\,,
\eear
with $\mathsf{H}$ the Hamiltonian. 
An interesting class of manifolds, where there is hope to find solutions of the e.o.m., is that of homogeneous spaces
\bea
\mathcal{M}={G\over H}\,,
\eea
where $G$ is a compact reductive group. In this case the space of homogeneous (i.e. $G$-invariant) metrics is finite dimensional~\cite{Kobayashi}, and its dimension $M$ depends on the particular manifold. An important subclass is that of symmetric spaces, where $M=1$, meaning that the metric is unique, up to an overall multiple.

In general, even for a homogeneous space the problem of finding geodesics in a generic invariant metric is not known to be integrable (cf.~\cite{Thimm, Alek}). However, there is a particular metric (called \emph{normal}), for which the problem is always soluble (cf.~\cite[Chapter X]{Kobayashi}). In the symplectic setup this metric may be described as follows (we will give an independent definition in section~\ref{reductmetrsect} below). On $\mathcal{M}$ one has the action of the Lie group $G$, which is then transferred to the action of $G$ on $T^\ast \mathcal{M}$. Accordingly, one can define a moment map\footnote{For the definition see~\cite[Part VIII]{ADS}. Intuitively, for any generator $T\in \mathsf{g}$ the function $\upmu(T)$ is the Hamiltonian for the action of the one-parametric subgroup $(e^{s T}, s\in \mathbb{R})$ on the manifold.}
\bea
\upmu:\quad T^\ast \mathcal{M}\quad \mapsto \quad \mathsf{g}^\vee\,,
\eea
where $\mathsf{g}$ is the Lie algebra of $G$. Since the Lie algebra $\mathsf{g}$ is reductive, one can pick a non-degenerate $ad$-invariant metric $\langle \bullet , \bullet \rangle$ on it. In practice, this metric will be the trace form $\langle a , b \rangle=-\mathrm{Tr}(ab)$ in a matrix representation of $\mathsf{g}$. Using this metric, one can identify $\mathsf{g}^\vee\simeq \mathsf{g}$, so that we will interchangeably view $\upmu$ as an element  of either $\mathsf{g}^\vee$ or~$\mathsf{g}$ (i.e., by a slight abuse of notation, $\upmu(T)\equiv \langle \upmu, T\rangle$ for any element $T\in \mathsf{g}$).

Given this setup, consider the Hamiltonian
\bea\label{Hmu2}
\mathsf{H}=-{1\over 2}\mathrm{Tr}\left(\upmu^2\right)
\eea
Let us show that the corresponding Hamilton's equations correspond to geodesic flow in a certain metric. Indeed, in this setup components of the moment map have the form
\bea
\upmu(T)=p_i v^i_{T}(x)\,,\quad\quad T\in \mathsf{g}\,.
\eea
where $T$ is a (fixed) generator in $\mathsf{g}$, and $v_{T}=v^i_{T}\,{\dd \over \dd x^i}$  is the vector field on $\mathcal{M}$ generating the flow of the one-parameter subgroup $\{e^{s T}\,,\; s\in\mathbb{R}\}\subset G$. The Hamiltonian may then be written as
\bea
\mathsf{H}={1\over 2}\sum\limits_{a}\,p_i p_j\,v^i_{T_a}\, v^j_{T_a}\,,
\eea
where the sum is over an orthonormal set of generators $T_a$ of $\mathsf{g}$. 
One thus sees that this is a special case of the Hamiltonian~(\ref{metrham}), where the inverse metric is
\bea\label{kappametr}
\vkappa^{ij}=\sum\limits_{a}\,v^i_{T_a}\, v^j_{T_a}
\eea
Such metrics have recently featured in~\cite{CY} in the context of 2D integrable models that we will discuss below.

Let us first explain, why one can solve Hamilton's equations in this case. The important observation is that, due to the $G$-symmetry of the problem, by Noether's theorem the moment map $\upmu$ (more exactly, the composition $\upmu\circ y: \mathbb{R}_t \mapsto \mathsf{g}$, which we denote by the same symbol) is conserved
\bea
\dot{\upmu}=0\quad\quad \Longrightarrow\quad\quad \upmu=\mathrm{const.}
\eea
E.o.m. obtained by varying w.r.t. $p$ are as follows:
\bea
\dot{x}=v_{\upmu}(x)
\eea
In other words, one has the flow lines of the vector field $v_\upmu$ corresponding to the one-parameter subgroup of $G$ generated by $\upmu$. We have thus fully solved the problem, the answer being that all geodesics are orbits of one-parameter subgroups of $G$.

\subsection{Normal metric.}\label{reductmetrsect}

Before proceeding further, let us give an independent definition of the normal metric on $\mathcal{M}=G/H$. Consider the standard decomposition of the Lie algebra $\mathsf{g}$:
\bea\label{hmdecomp}
\mathsf{g}=\mathsf{h}\oplus \mathsf{m}\,,
\eea
where $\mathsf{m}$ is the orthogonal complement to $\mathsf{h}$ in the metric $\langle \bullet, \bullet \rangle$. Next, consider the Maurer-Cartan current $\hat{j}=-g^{-1}dg$\;($g\in G$), which is a one-form on $G$ with values in $\mathsf{g}$. More invariantly,
\bea\label{jonvect}
\hat{j}(v_{T})=-g^{-1}Tg\quad\quad \textrm{for any generator}\quad\quad T\in \mathsf{g}\,.
\eea
The conventional procedure is then to decompose $\hat{j}$ according to~(\ref{hmdecomp}):
\bea
\hat{j}=\hat{j}_{\mathsf{h}}+\hat{j}_{\mathsf{m}}
\eea
W.r.t. the action of $H$, $g\to gh$ with $h\in H$, one has $\hat{j}_{\mathsf{m}} \to h^{-1} \hat{j}_{\mathsf{m}} h$. It is more convenient to deal with the gauge-invariant one-form $\hat{J}_{\mathsf{m}}:=g\hat{j}_{\mathsf{m}} g^{-1}$. One can therefore build a metric with the line element
\bea
ds^2=\langle\hat{J}_{\mathsf{m}}, \hat{J}_{\mathsf{m}}\rangle\,.
\eea
This is what is known as the normal metric $\vkappa_{\mathrm{norm}}$. Let us show that it matches the previous definition, so that $\vkappa_{\mathrm{norm}}=\vkappa$. Since~(\ref{kappametr}) defines the inverse metric $\vkappa^{-1}$, it will be convenient to deal with one-forms rather than vector fields. Recall that on $\mathcal{M}$ one has the basic vector fields $v_{T_a}$ and also the one-forms
\bea
\theta_{T_a}=\langle T_a, \hat{J}_{\mathsf{m}} \rangle\quad  \textrm{for}\quad T_a\in \mathsf{g}\,,
\eea
which are related by the metric~(\ref{kappametr}) as follows:
\bear
&&\vkappa^{-1}(\bullet, \theta_{T_b})=\sum_{a} v_{T_a}\, \langle T_a, \hat{J}_{\mathsf{m}}(v_{T_b})\rangle\equiv v_{\hat{J}_{\mathsf{m}}(v_{T_b})}:=\theta_{T_b}^{\vee}\,,\\
&& \textrm{where}\quad\quad \hat{J}_{\mathsf{m}}(v_{T_b})=-g[g^{-1}T_bg]_{\mathsf{m}} g^{-1}\,.
\eear
In computing the expression in the last line, we used~(\ref{jonvect}). Next, notice that
\bea
\vkappa_{\mathrm{norm}}(\theta_{T_b}^{\vee}, \theta_{T_c}^{\vee})=\langle[g^{-1}T_b g]_{\mathsf{m}}, [g^{-1}T_cg]_{\mathsf{m}}\rangle\,.
\eea
On the other hand,
\bea
\vkappa(\theta_{T_b}^{\vee}, \theta_{T_c}^{\vee})=\theta_{T_b}(\theta_{T_c}^{\vee})=\langle T_b, g[g^{-1}T_cg]_{\mathsf{m}}g^{-1}\rangle=\vkappa_{\mathrm{norm}}(\theta_{T_b}^{\vee}, \theta_{T_c}^{\vee})\,,
\eea
which proves that the two metrics are the same.

\subsection{Quantum mechanics.}

Another important point is that, for this metric, even the quantum problem can be explicitly solved. Upon quantization, Hamiltonian~(\ref{Hmu2}) is mapped to $\hat{\mathsf{H}}=-\triangle$, where $\triangle$ is the Laplacian in the normal metric. We will not prove it here (cf.~\cite[Appendix B.4]{Camporesi}), but one can as well interpret the Hamiltonian
\bea
\hat{\mathsf{H}}=\mathcal{C}_2
\eea
as the quadratic Casimir operator acting in the Hilbert space
\bea\label{Hdecomp}
\mathscr{H}=L^2\left(G/H\right)=\oplus_{i\in \mathcal{I}} \,V_i\otimes \CC^{m_i}\,,
\eea
where the index runs over a subset $\mathcal{I}$  of the set of irreducible representations $V_i$ of~$G$, and $m_i$ is the multiplicity of the representation. The decomposition~(\ref{Hdecomp}) can be constructed by means of the Peter-Weyl theorem\footnote{The theorem of Peter and Weyl provides an explicit way of decomposing $L^2\left(G/H\right)$ into irreducible representations of $G$ (generalized spherical harmonics): $L^2\left(G/H\right)=\oplus_{\textrm{irreps}\;\textrm{of}\;G} \,V_i\otimes W_i^\ast$, where the sum is over all unitary irreps of $G$, and $W_i^\ast\subset V_i^\ast$ is the subspace of $V_i^\ast$, on which $H$ acts trivially. For more on this cf.~\cite{Segal}.}. This means that the problem of finding the spectrum of $\hat{\mathsf{H}}$ is reduced to the problem of finding the values of the Casimir $\mathcal{C}_2$ in the representations $V_i$.

\subsubsection{The spheres.}\label{spheresect} 
As an illustration, where the Hilbert space and spectrum may be written out explicitly, consider the case of a sphere $G/H=S^{n-1}$. It will be beneficial to represent the sphere as a quotient
\bea
S^{n-1}=\left(\mathbb{R}^n\setminus\{0\}\right)/\mathbb{R}^+\,,
\eea
where $\mathbb{R}^+$ is the multiplicative group of positive real numbers. The reason is that the symmetry group $G=O(n)$ acts linearly in $\mathbb{R}^n$, so that the vector fields $v_{T_a}(x)$ are linear. First, consider the free action
\bea
\mathcal{S}_0=\int\,dt\,\left(\sum_{i=1}^n\,p_i (\dot{x}^i-A x^i)\right)\,,
\eea
where $x^i$ are the coordinates in $\mathbb{R}^n$, and the quotient by $\mathbb{R}^+$ is realized through a gauge field implementing symplectic reduction in the Hamiltonian framework. Even though this formulation may seem redundant at this stage, it ultimately simplifies the framework tremendously. Gauge transformations act as follows:
\bea
x^i\mapsto \lambda x^i,\;p_i\mapsto \lambda^{-1} p_i,\; A\mapsto A+{d\over dt}\left(\log{ \lambda}\right)\,,\quad\quad \lambda \in \mathbb{R}^+\,.
\eea
This system describes the symplectic reduction
\bea
\left(T^\ast \mathbb{R}^n\setminus\{0\}\right)\,/\!/\,\mathbb{R}^+\simeq \Bigl\{\sum_i\,p_i x^i=0\subset T^\ast \mathbb{R}^n\setminus\{0\}\Bigl\}\,/\,\mathbb{R}^+\,,
\eea
where the expression in brackets is the moment map constraint obtained by varying w.r.t. the gauge field. The moment map for the action of the global symmetry group $O(n)$ is $\upmu=x\otimes p-p\otimes x\in \mathsf{o}(n)$. Upon adding the Hamiltonian~(\ref{Hmu2}), we get
\bear\label{mechact}
&&\mathcal{S}=\mathcal{S}_0+{1\over 2}\,\int\,dt\,\mathrm{Tr}(\upmu^2)\,=\\ \nonumber
&&= \int\,dt\,\left(\sum_i\,p_i (\dot{x}^i-A x^i)-\sum_j (x^j)^2\cdot \sum_k (p_k)^2+\left(\sum_i x^i p_i\right)^2\right)\,.
\eear
Notice that the last term is proportional to the constraint obtained by varying w.r.t.~$A$, so that it may be dropped. Alternatively, it may be eliminated by the shift $A \mapsto A+ \sum_i x^i p_i$.

The system~(\ref{mechact}) may be readily quantized. We postulate the canonical commutation relations $[x^j, p_k]=i\,\delta_k^j$, so that $ p_k=-i{\dd \over \dd x^k}$. The Hamiltonian and the constraint may be read off from~(\ref{mechact}):
\bea\label{hamconstr}
\hat{\mathsf{H}}\psi=-\sum_j (x^j)^2\cdot \sum_k {\dd^2 \psi \over \dd (x^k)^2}\,,\quad\quad \mathcal{C}\psi=\sum_j x^j {\dd \psi\over \dd x^j}=0\,.
\eea
The constraint implies that $\psi$ should be a function of the ratios $x_i\over x_j$ and, as such, is a well-defined function on the quotient $\left(\mathbb{R}^n\setminus\{0\}\right)/\mathbb{R}^+$. This is the reason why we have chosen the particular operator ordering in the constraint in~(\ref{hamconstr}). For eigenfunctions of the Hamiltonian one can choose a natural ansatz of homogeneous functions (here $ \|x\|^2=\sum_i\,(x^i)^2$):
\bea
\psi_{\ell}={1\over \|x\|^{\ell}}\,\sum\limits_{i_1, \ldots, i_{\ell}=1}^n\,A_{i_1\cdots i_{\ell}}\,x^{i_1}\cdots x^{i_{\ell}}\,,
\eea
with $A_{i_1\cdots i_{\ell}}$ constant. One easily checks that, in order for this to be an eigenfunction, $A$ should be traceless: $\sum_j\,A_{j\,j\,i_3\cdots i_{\ell}}=0$ for any $i_3, \ldots, i_{\ell}$. In other words, the expression in the numerator is a harmonic polynomial, whereas the denominator effectively means we should restrict it to the sphere $S^{n-1}$. In this way we recover the familiar relation of spherical harmonics with harmonic polynomials. Acting on $\psi_{\ell}$ with $\hat{\mathsf{H}}$, we get the spectrum:
\bea
\hat{\mathsf{H}}\psi_{\ell}=\ell(\ell+n-2)\psi_{\ell}\,,\quad\quad\ell=0, 1, 2, \ldots
\eea
Note also that, by restricting to even $\ell=0, 2, 4, \ldots$, we obtain the spectrum of the real projective space $\mathbb{RP}^{n-1}$. Indeed, here one has a quotient by $\mathbb{R}^\ast$ rather than by $\mathbb{R}^+$, and only the states $\psi_{\ell}$ with even $\ell$ are invariant.

A similar construction works in the case of $\CP^{n-1}$ and presumably also in the case of the quaternionic projective space $\mathbb{HP}^{n-1}$. All relevant spectra can be found in the book~\cite{Berger}.

\section{Reductions of the principal chiral model}\label{PCMsect}

The mechanical problem of geodesic motion has a natural 2D generalization, known as the theory of harmonic maps, or sigma models in physics language. These are maps $x: \Sigma \to \mathcal{M}$ from a Riemann surface~$\Sigma$, which are critical points of the action\footnote{Note that classically the sigma model action depends only on the conformal class of the worldsheet metric due to the so-called Weyl invariance.  If one brings the metric to conformal coordinates, $ds^2=e^{\Lambda}\,dz d\bar{z}$, the factor $e^{\Lambda}$ drops out of the action, and the remaining degrees of freedom are encoded in the complex coordinates $z, \bar{z}$. By varying the complex structure on $\Sigma$, one effectively varies the metric. As a result,~(\ref{2daction}) takes into account the choice of metric via the choice of complex coordinates.}
\bea\label{2daction}
\mathcal{S}_{2D}=\int_{\Sigma}\,d^2z\,{1\over 2}\,\vkappa_{ij}\,\dd x^i\, \thickbar{\dd}x^j
\eea
We will start with the case when $\mathcal{M}={G}$ is a group, i.e. the principal chiral model (PCM). As before, we have the Maurer-Cartan current $\hat{j}=-g^{-1}dg\in \Lambda^1(G)\otimes \mathsf{g}$ (where again $g\in G$ is the group element). Accordingly, we denote by $j=x^\ast \hat{j}$ its pull-back to~$\Sigma$. The defining equations of the PCM are
\bear
&&dj-j\wedge j=0\\
&& d\ast j=0\,.
\eear
Here $\ast$ is the Hodge star, whose action on one-forms coincides with the complex structure on~$\Sigma$ (by the same reason mentioned in the footnote). It acts as $\ast dz=i dz$, $\ast \thickbar{dz}=-i\,\thickbar{dz}$. One may decompose $j$ in these basic one-forms: $j=i\,(j_z dz+\thickbar{j_z} \thickbar{dz})$. The equations may then be rewritten as a single complex equation
\bea\label{PCMeq}
\thickbar{\dd} j_z+{i\over 2}\,[j_z, \thickbar{j_z}]=0\,.
\eea
Now, take $j_z$ in the standard representation of the corresponding Lie algebra (for example, as an $n\times n$ matrix for $G=U(n)$). It follows from~(\ref{PCMeq}) that $\thickbar{\dd} \mathrm{Tr}\left(j_z^\ell\right)=0$ for any $\ell=1, 2, 3, \ldots $, so that $a_{\ell}(z):=\mathrm{Tr}\left(j_z^\ell\right) dz^{\ell}$ are holomorphic sections of $K_{\Sigma}^{\otimes \ell}$ -- a power of the canonical bundle.  
The reduction that we wish to perform amounts to fixing these sections $a_{\ell}(z)$. Consider the following special cases:

\begin{itemize}
\item $\Sigma=\CP^1$. Here $K_{\Sigma}=\mathcal{O}(-2)$ (square of the tautological bundle), and there are no non-zero holomorphic sections of $K_{\Sigma}^{\otimes \ell}$ with positive $\ell$. It then follows that $j_z^m=0$ for some positive integer $m$.

\item $\Sigma=\CC$, with the additional condition of `Lorentz' (rotational) invariance. Since $j_z$ transforms as $j_z \mapsto e^{i\theta} j_z$, any non-zero fixed $a_{\ell}(z)$ would violate Lorentz invariance. As a result, again we have $j_z^m=0$.

\item $\Sigma=\mathbb{T}^2$, the 2-torus. Here the canonical class is trivial, so that $a_{\ell}$ are constant.
\end{itemize}

Here we will only deal with the case that $j_z$ is nilpotent, $j_z^m=0$. In a neighborhood\footnote{By neighborhood we mean an $\epsilon$-disc $\|z-z_0\|<\epsilon$, where distance is measured w.r.t. the induced metric $\left(ds^2\right)_{\Sigma}=\mathrm{Tr}(j_z \thickbar{j_z}) dz \thickbar{dz}$.} of every point one can write the solution to~(\ref{PCMeq}) as follows:
\bea
j_z=\hat{g}(z, \thickbar{z})Q(z)\hat{g}^{-1}(z, \thickbar{z})\,,
\eea
with $Q(z)$ holomorphic. Thus, the Jordan type of $j_z$ may only change in isolated points on $\Sigma$, where $Q(z)$ degenerates. We may thus assume that, in a generic point on $\Sigma$, $j_z\in N\subset \mathsf{g_C}$, where $N$ is a nilpotent orbit\footnote{Let $x\in \mathsf{g_C}$ be an $ad$-nilpotent element in the Lie algebra (i.e. $(ad_x)^m=0$ for some $m$). By the nilpotent orbit $N_x$  one means the adjoint orbit of $x$: $N_x=\{\,g x g^{-1}\,,\; g\in G_{\CC}\}$. One can show that in this case $x$ is a nilpotent matrix in any representation of $\mathsf{g_C}$, cf.~\cite{Collingwood}. In practice, when speaking of Jordan forms, we will always assume we are dealing with $x$ in the standard (defining) representation of the corresponding Lie algebra.} of $G_{\CC}$. As we shall see, calculations are greatly simplified by a proper choice of coordinates on $N$ (and on its closure $\thickbar{N}$).  

Let us consider two examples, which share some common features but also have important distinctions: these are the minimal orbits in $\mathsf{sl_n}$ and in $\mathsf{sp_{2n}}$. Both of these may be described as sets of non-zero matrices $j_z$ in the corresponding Lie algebras, satisfying $j_z^2=0$ and being of rank one. Without loss of generality, they may be parametrized as follows:
\begin{itemize}
\item $\mathsf{sl_n}$:\quad let $U, V\in \CC^n$, then  $j_z=U\otimes V$, with $VU=0$.
\item $\mathsf{sp_{2n}}$: let $W\in \CC^{2n}$ and $\omega_{2n}$ the non-degenerate skew-symmetric form, then $j_z=W\otimes W^t \omega_{2n}$.
\end{itemize}

Substituting these parametrizations in the PCM equation~(\ref{PCMeq}), we obtain the following equations for $U, V, W$ (and their complex conjugates):
\bear\label{UVeq}
&&\thickbar{D} U-{i\over 2}\, \left(\thickbar{U}U\right) \thickbar{V}=0\,,\quad\quad \thickbar{D}V+{i\over 2}\, \left(V\thickbar{V}\right) \thickbar{U}=0\,.\\
&& \label{Weq} \thickbar{\dd} W+{i\over 2}\,\omega_{2n}\left(\thickbar{W}W\right)\,W^\ast=0\,.
\eear
In the first line, $\thickbar{D} U:=\thickbar{\dd} U-i \thickbar{\mathcal{A}}U$ and $\thickbar{D} V:=\thickbar{\dd} V+i \thickbar{\mathcal{A}}V$, where to simplify the notation we have introduced an auxiliary $\CC^\ast$ gauge field\footnote{One can find $\thickbar{\mathcal{A}}$, for example, by taking the scalar product of the first equation in~(\ref{UVeq}) with $\thickbar{U}$ and using the constraint $VU=0$.} $\thickbar{\mathcal{A}}$, the $\CC^\ast$ transformation properties of the fields being  $U\to e^{\chi} U\,,\;V\to e^{-\chi} V$. We might express $\thickbar{V}$ from the first equation, so that, substituting into the second one, we obtain 
\bea\label{CPNeq}
\thickbar{D}D\thickbar{U}+\uprho\, \thickbar{U}=0\,,\quad\quad \uprho=\frac{D\thickbar{U}\thickbar{D}U}{\thickbar{U}U}\,.
\eea
which this the e.o.m. of the $\CP^{n-1}$ sigma model. On the other hand, in general there is no analogous transformation that one could perform on equation~(\ref{Weq}) to bring it to second-order form, thus in that case there is no equivalence with sigma models. As we shall see in section~\ref{gensec}, the first case corresponds to the \emph{chiral} GN-model, whereas the second one is the \emph{non-chiral} GN-model.

Motivated by these two examples, we may ask which nilpotent orbits lead to sigma models. Conjecturally, the answer is that these are orbits $N$ whose closure $\thickbar{N}$ admits a symplectic resolution of singularities ($\thickbar{N}$ is a singular variety). In this case the resolved space is $T^\ast\mathcal{F}$~\cite{Fu}, the cotangent bundle of some flag manifold. The latter is the target space of the sigma model. In the next section we recall the concept of such resolutions of singularities\footnote{For an introductory exposition of some of the material in this section see~\cite{Ginzburg}. More advanced topics on symplectic singularities and their resolutions are covered in~\cite{Fu, Namikawa}.}. 
 
\section{The Springer resolution}

In contrast to the examples of the previous section, we start with the generic (i.e., in a sense polar to the minimal one) nilpotent orbit $N_{\mathrm{reg}}$, called regular. Consider the quotient
\bea\label{Vdef}
V:= G_{\CC}\times n(\mathsf{b}) / B\equiv G_{\CC} \times_B n(\mathsf{b})\,,
\eea
where $B\subset G_{\CC}$ is the Borel subgroup, $\mathsf{b}$ its Lie algebra, and $n(\mathsf{b})$ its nilradical\footnote{In the case $G_{\CC}=SL(n, \CC)$ the Borel subgroup consists of invertible upper-triangular matrices, whereas $n(\mathsf{b})$ comprises  strictly upper-triangular matrices.}. The quotient is taken w.r.t. the following action of $B$:
\bea\label{homaction}
(g,\;\beta)\quad \sim \quad (gh,\; h^{-1}\beta h)\,,\quad\quad g\in G_{\CC}\,,\; \beta\in n(\mathsf{b})\,,\; h\in B\,.
\eea

Let us show that $V\simeq T^\ast \left(G_{\CC}/B\right)$, where $G_{\CC}/B$ is the manifold of complete flags. To prove this, we use the fact that on any homogeneous space $G/H$ the tangent bundle may be expressed as\footnote{Indeed, one has a map
\bea\label{tanspacemap}
\left(G\times \mathsf{g}\right)/H \mapsto T\left(G/H\right)
\eea
constructed as follows. Let $(g, a)\in \left(G\times \mathsf{g}\right)/H$ and $f(gH)=f(g)$ an arbitrary function on $G/H$. We may then define a vector field $v$ on $G/H$ by $v f(gH):={d\over d\epsilon}\,f\left(g\,e^{\epsilon a}H\right)\big|_{\epsilon=0}$. Since $G/H$ is homogeneous, all vector fields are constructed this way, so the map~(\ref{tanspacemap}) is surjective. Besides, two elements $(g, a_1)$ and $(g, a_2)$ are mapped to the same vector fields if and only if $a_2-a_1\in \mathsf{h}$. Thus, replacing $\mathsf{g}$ in~(\ref{tanspacemap}) by the quotient $\mathsf{g}/\mathsf{h}$ makes the map one-to-one, and one arrives at~(\ref{tanhomspace}).}
\bea\label{tanhomspace}
T\left(G/H\right)\simeq \left(G\times \mathsf{g}/\mathsf{h}\right)/H\,,
\eea
where the action of $H$ is $(g, a)\to (gh, h^{-1}ah)$. To pass to the cotangent bundle, one should replace $\mathsf{g}/\mathsf{h}$ in~(\ref{tanhomspace}) with its dual $\left(\mathsf{g}/\mathsf{h}\right)^\vee$. The action of $H$ on $\mathsf{g}/\mathsf{h}$ is then inherited by the dual space: if $a\in \mathsf{g}/\mathsf{h}$ and $\beta \in \left(\mathsf{g}/\mathsf{h}\right)^\vee$, one has $[h\circ\beta](h\circ a)\equiv \beta (a)$. In our case $\mathsf{h}=\mathsf{b}$, and $\left(\mathsf{g}/\mathsf{b}\right)^\vee\simeq n(\mathsf{b})$. Viewing $n(\mathsf{b})$ as the space of strictly upper-triangular matrices, the pairing is simply given by the trace form: $\beta(a)=\mathrm{Tr}(\beta a)$. This is well-defined on the quotient $\mathsf{g}/\mathsf{b}$ since for $a_0\in \mathsf{b}$ one has $\mathrm{Tr}(\beta a_0)=0$.  The action of $H=B$ on $n(\mathsf{b})$ is $h\circ \beta=h^{-1}\beta h$, so that one arrives at the definition~(\ref{homaction}).

We can construct a canonical one-form on the cotangent bundle $V$ as follows:
\bea
\theta=\mathrm{Tr}\left(\beta g^{-1}dg\right)
\eea
To check that this is a well-defined form on the quotient, one makes the transformation $g\to gh$ with $h\in B$:
\bea\label{gaugetrans1}
\theta \mapsto \theta+\mathrm{Tr}\left(\beta dh h^{-1}\right)=\theta\,,
\eea
since $\beta \in n(\mathsf{b})$ and $dh h^{-1}\in \mathsf{b}$. 
Given the above identification with the cotangent bundle, $\theta$ may be viewed as the canonical one-form. Its derivative is the symplectic form on the cotangent bundle:
\bea
\Omega=d\theta
\eea

We can as well define the projection map
\bea\label{SpringerProj}
\pi_{\textrm{Spring}}:\quad V \mapsto \thickbar{N}_{\mathrm{reg}}\,,\quad\quad \pi_{\textrm{Spring}}(g, \beta)=g\beta g^{-1}
\eea
known as the Springer resolution. One can show that $\upmu=g\beta g^{-1}$ is the moment map corresponding to the action of~$G_{\CC}$ on $V$.

To show that $\pi_{\textrm{Spring}}$ is a resolution of singularities, first of all recall that $n(\mathsf{b})\subset \thickbar{N}_{\mathrm{reg}}$. Consider $n_0(\mathsf{b}):=n(\mathsf{b}) \bigcap N_{\mathrm{reg}}$, i.e. the set of `generic' elements of $n(\mathsf{b})$. One can show that $B$ acts transitively on $n_0(\mathsf{b})$, implying that it can bring any element to the reference one: $h_0^{-1} \beta h_{0}=\beta_0$ for $\beta\in n_0(\mathsf{b})$ and $h_0\in B$. As a result,
\bea
\pi_{\textrm{Spring}}^{-1}(N_{\mathrm{reg}})=G_{\CC}\times n_0(\mathsf{b}) / B\simeq G_{\CC}/ G_{\mathrm{Stab}}\,,
\eea
where $G_{\mathrm{Stab}}\subset B$ is the stabilizer of an element $\beta_0\in n_0(\mathsf{b})$. Since $G_{\CC}/ G_{\mathrm{Stab}}\simeq N_{\mathrm{reg}}$, one finds that $\pi_{\textrm{Spring}}^{-1}(N_{\mathrm{reg}})\simeq N_{\mathrm{reg}}$, so that $\pi_{\textrm{Spring}}$ is a one-to-one map outside of the singular set.

Pick an element $\beta_0\in n_0(\mathsf{b})$. The one-form
\bea\label{theta0}
\theta_0=\mathrm{Tr}\left(\beta_0 g^{-1}dg\right)
\eea
is a well-defined one-form on $G_{\CC}/ G_{\mathrm{Stab}}$. This can be established by checking invariance w.r.t. a gauge transformation $g\to gk$ with $k\in G_{\mathrm{Stab}}$, as in~(\ref{gaugetrans1}) above. Its derivative
\bea
\Omega_0=d\theta_0=-\mathrm{Tr}\left(\beta_0 g^{-1}dg\wedge g^{-1}dg\right)
\eea
is the Kirillov-Kostant symplectic form on the nilpotent orbit $N_{\mathrm{reg}}$. Notice that it is exact, unlike the analogous two-form on the adjoint orbits of compact groups (the flag manifolds themselves).

\subsection{Example: the nilpotent orbit of $GL(2, \CC)$.}\label{sl2example} Consider the case $G_{\CC}=GL(2, \CC)$, and pick $\beta_0=\begin{pmatrix} 0 &1 \\  0 & 0\end{pmatrix}$. In this case $G_{\mathrm{Stab}}=\Biggl\{ \begin{pmatrix} \alpha &\gamma \\  0 & \alpha\end{pmatrix} \Biggl\} \subset B$. One can parametrize an element of the quotient $N=G_{\CC}/G_{\mathrm{Stab}}$ as follows:
\bea
g=\begin{pmatrix} U_1 &\bullet \\  U_2 & \bullet\end{pmatrix}\,,\quad\quad g^{-1}=\begin{pmatrix} \bullet &\bullet \\  V_1 & V_2\end{pmatrix}
\eea
Since $gg^{-1}=\mathds{1}_2$, one has $\sum_{i=1}^2 \,V_iU_i=0$. Besides, one has a residual quotient $U_i\to \alpha U_i$, $V_i\to \alpha^{-1}V_i$ with $\alpha \in \CC^\ast$. In these variables, the one-form~(\ref{theta0}) is written as
\bea
\theta_0=\sum_{i=1}^2 \,V_i\,dU_i
\eea
This is the canonical one-form on $N$, where $N$ (the open part of $\thickbar{N}$) is defined by the conditions $(U_1, U_2)\neq 0$ and $(V_1, V_2)\neq 0$. It is instructive to consider the corresponding moment map
\bea
\upmu:=g \beta_0 g^{-1}=\begin{pmatrix} U_1 \\ U_2\end{pmatrix} \otimes \begin{pmatrix} V_1 & V_2\end{pmatrix}\in \mathsf{sl}(2, \CC)
\eea
On $N$, $\upmu\neq 0$, whereas the closure corresponds to adding the point $\upmu=0$. The $(U, V)$-variables are the same ones that featured in~(\ref{UVeq}), since in the special case of $GL(2, \CC)$ the minimal orbit is the same as the regular one.

\subsection{Grassmannians and further generalizations.}\label{gensec}

As a generalization, we can now replace in~(\ref{Vdef}) the Borel subgroup $B$ with an arbitrary parabolic subgroup $P$ of $G_{\CC}$, and accordingly $\mathsf{b}$ with the Lie algebra $\mathsf{p}$ of $P$. The image of~(\ref{SpringerProj}) will then be the closure of a certain nilpotent orbit. In general, though, not all nilpotent orbits may be obtained in this way.  The ones that can be are called~\emph{Richardson} orbits. Conjecturally these are the ones that are of interest for our sigma model applications\footnote{For such orbits there is also a canonical choice of $(U, V)$-type Darboux coordinates, i.e. a polarization, as shown in~\cite{KamalinP}.}.

A simple example, in a sense polar to the one of the complete flag manifold, is that of a Grassmannian $Gr(k, n)$. Let us assume that $n-k\geq {n\over 2}$, or else we switch $k \to n-k$. The corresponding parabolic subgroup is shown in Fig.~1 (a).
\begin{figure}
\begin{center}
\begin{minipage}{.2\textwidth}
 \begin{tikzpicture}[every node/.style={text height=1ex,text width=0.5em}]
 \draw[<->, line width=0.3mm] (65pt,13pt) to (65pt,40pt);
 \put(73,23){$k$};
  \put(-5,-60){$(a)$};
  \put(200,-60){$(b)$};
\matrix[matrix of math nodes,
        left delimiter=(,
        right delimiter=),
        nodes in empty cells] (m)
{
       &  &{}  &  &  &  \\
  &         &   &{}   &       &   \\
       &   &   &   &       &   \\
       &   &   &   &       &   \\
  &   &   &   &       &   \\
       &   &   &   &       &   \\
};

\def\mypath{(m-1-1.north west) node[left] {}
    -- (m-1-6.north east) node[below] {}
    -- (m-6-6.south east) node[below] {}
    -- (m-6-3.south west) node[below] {}
    -- (m-2-3.south west) node[below] {}
    -- (m-2-1.south west) node[below] {} 
    -- cycle;
    };
    \path[draw, line width = 0.3mm] (m-1-3.north west) -- (m-6-3.south west);
\path[draw, line width = 0.3mm] (m-3-1.north west) -- (m-3-6.north east);
    \draw [pattern={Lines[distance=1mm,angle=45,line width=0.3mm]},
        pattern color=grey] \mypath;
    \end{tikzpicture}
    \end{minipage} 
    \hspace{4cm}
    \begin{minipage}{.2\textwidth}
     \begin{tikzpicture}[every node/.style={text height=1ex,text width=0.5em}]
\put(-75,-7) {$\beta_0=$};
\matrix[matrix of math nodes,
        left delimiter=(,
        right delimiter=),
        nodes in empty cells] (m)
{
       &  & 1  &  &  &  \\
  &         &   & 1  &       &   \\
       &   &   &   &       &   \\
       &   &   &   &       &   \\
  &   &   &   &       &   \\
       &   &   &   &       &   \\
};
\path[draw, line width = 0.3mm] (m-1-3.north west) -- (m-6-3.south west);
\path[draw, line width = 0.3mm] (m-3-1.north west) -- (m-3-6.north east);
    \end{tikzpicture}
    \end{minipage}
    \end{center}
    \caption{$(a)$: parabolic subgroup $P$, corresponding to Grassmannian $Gr(k, n)$. \\$(b)$: typical element in $n(\mathsf{p})\bigcap N$ (here $k=2$).}
    \end{figure}
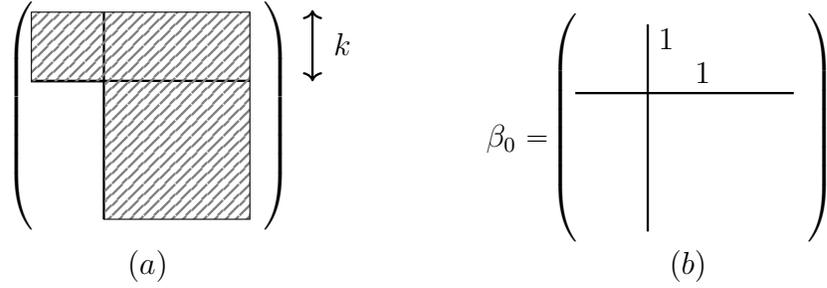

The resolution of singularities has the form
\bea
T^\ast Gr(k, n) \mapsto \thickbar{N}\,,
\eea
where the typical element of the nilpotent orbit $N$ is shown in Fig.~1 (b), where we have chosen $k=2$ for concreteness. Notice that we have chosen $\beta_0$ to lie in $n(\mathsf{p})\bigcap N$, as described earlier. For general $k$, the corresponding Jordan structure is of the type $(2^k, 1^{n-2k})$.

As already mentioned, this algorithm may be applied to various parabolic subgroups\footnote{As discussed in~\cite{Namikawa}, for an arbitrary parabolic subgroup the relevant analogue of the Springer map~(\ref{SpringerProj}) is a surjective map of degree $d\geq 1$ and only defines a resolution of singularities whenever $d=1$. At present it is unclear what the implications of $d>1$ are for the relation to sigma models. In the foregoing we thus restrict to those cases when $d=1$.}. Let us summarize our discussion by the following commutative diagram:

\vspace{0.3cm}
\begin{center}
 \begin{tikzpicture}
 \put(0,0){$\Sigma$} ;
 \put(50,0){$\thickbar{N}$};
 \put(48,50){$T^\ast \mathcal{F}$};
 \put(0,50){$\mathcal{F}$} ;
 \put(50,25){};
  \draw[->, line width=0.3mm] (40pt,52pt) to (15pt,52pt);
    \draw[->, line width=0.3mm] (15pt,02pt) to (45pt,02pt);
    \draw[->, line width=0.3mm] (15pt,15pt) to (45pt,45pt);
    \draw[->, line width=0.3mm] (3pt,15pt) to (3pt,45pt);
    \draw[->, line width=0.3mm] (55pt,45pt) to (55pt,15pt);
    \put(25, -7){$j_z$};
    \put(25, 57){$\pi$};
    \put(60, 28){$\pi_{\mathrm{Spring}}$};
    \node[rotate=45] at (25pt,35pt) {$\mathrm{lift}$};
  \end{tikzpicture}
    \end{center}
    Here $j_z$ is the map to nilpotent matrices given by the $z$-component of the current, as described in section~\ref{PCMsect}, $\pi_{\textrm{Spring}}$ is the Springer resolution, $\pi$ is the standard fiber bundle projection and `$\mathrm{lift}$' is the lift of the map from $\Sigma$ to the resolution. The existence of this lift is highly non-trivial and implies that the e.o.m. may be split in two sets of equation, one for $U$ and another one for $V$, as it happens in~(\ref{UVeq}), (\ref{CPNeq}). The composition $\pi \circ \mathrm{lift}: \Sigma \to \mathcal{F}$ is the map to the target space of the resulting sigma model\footnote{We have shown in~\cite{Bykov1} that these maps are well-defined in the case of $\mathsf{sl_n}$-flags, where the relevant flag is the flag of kernels $\mathrm{Ker}(j_z^\ell), \;\ell=1, 2, \ldots$}.

\subsection{The quiver representation.}

Let us now explain that on $\thickbar{N}$ there is a preferred choice of coordinates, which is related to the fact that $\thickbar{N}$ is a quiver variety\footnote{In the $\mathsf{sl}$ case this was shown in~\cite{Nakajima}, and generalized to $\mathsf{o}$- and $\mathsf{sp}$-orbits in~\cite{Kobak}.}. We will adopt the following working definition of quiver variety:
\bea
Q:= \Phi_0 // \hat{G}_{\CC}\,,\quad\quad \Phi_0\simeq \CC^{2N}\,.
\eea
This means that $Q$ is a complex symplectic quotient of flat space $\CC^{2N}$ by a complex reductive group $\hat{G}_{\CC}$. Typically symplectic reduction deals with the action of a group preserving the symplectic form, i.e. $\pounds_{\xi} \Omega=0$. However, in all of our examples a stronger condition holds: there is a canonical one-form $\uptheta$ such that $\Omega=d\uptheta$, and all relevant symmetries preserve $\uptheta$, i.e. $\pounds_{\xi} \uptheta=0$. 

The space $\CC^{2N}$ and the action of the group on it may be conveniently encoded by a quiver. For example, the quivers for the minimal orbits are as follows:

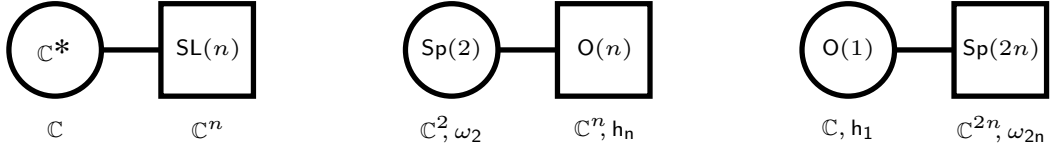
\vspace{0.5cm}
\begin{tikzpicture}
\tikzstyle{every node}=[font=\tiny];
\draw[line width=2pt] (0,0) circle (17.5pt);
\draw[line width=2pt] (17.5pt,0pt) -- (40pt,0);
\draw[line width=2pt] (40pt,-17.5pt) rectangle ++(35pt,35pt);
\node[scale=1.4] at (0pt,-30pt) {$\CC$};
\node[scale=1.4] at (0pt,0pt) {$\CC^\ast$};
\node[scale=1.4] at (58pt,0pt) {$\mathsf{SL}(n)$};
\node[scale=1.4] at (58pt,-30pt) {$\CC^n$};

\draw[line width=2pt] (150pt,0) circle (17.5pt);
\draw[line width=2pt] (167.5pt,0pt) -- (190pt,0);
\draw[line width=2pt] (190pt,-17.5pt) rectangle ++(35pt,35pt);
\node[scale=1.4] at (150pt,-30pt) {$\CC^2\!\!, \mathsf{\omega_2}$};
\node[scale=1.4] at (150pt,0pt) {$\mathsf{Sp}(2)$};
\node[scale=1.4] at (208pt,0pt) {$\mathsf{O}(n)$};
\node[scale=1.4] at (208pt,-30pt) {$\CC^n\!\!, \mathsf{h_n}$};

\draw[line width=2pt] (300pt,0) circle (17.5pt);
\draw[line width=2pt] (317.5pt,0pt) -- (340pt,0);
\draw[line width=2pt] (340pt,-17.5pt) rectangle ++(35pt,35pt);
\node[scale=1.4] at (300pt,-30pt) {$\CC, \mathsf{h_1}$};
\node[scale=1.4] at (300pt,0pt) {$\mathsf{O}(1)$};
\node[scale=1.4] at (358pt,0pt) {$\mathsf{Sp}(2n)$};
\node[scale=1.4] at (358pt,-30pt) {$\CC^{2n}\!\!, \mathsf{\omega_{2n}}$};
\end{tikzpicture}

Here $\mathsf{h_n}$ and $\mathsf{\omega_{2n}}$ are the symmetric and skew-symmetric forms on the respective spaces ($\mathsf{h_1}$ and $\mathsf{\omega_2}$ are unique up to a multiple). The line between the nodes means we have a minimal set of maps between the nodes that admits a symplectic structure. These are summarized in the table:

\vspace{0.3cm}
\begin{table}[H]
\begin{center}
\begin{tabular}{|c  c c |} 
 \hline
 Type & Matter content & Symplectic form   \\ [0.5ex] 
 \hline\hline
$\mathsf{sl_n}$ & $(U, V)\in\mathrm{Hom}(\CC, \CC^n)\oplus \mathrm{Hom}(\CC^n, \CC)$ & $dV\wedge  dU$  \\ 
 \hline
$\mathsf{o_n}$ &  $W\in \mathrm{Hom}(\CC^2, \CC^n)$ & $\mathrm{Tr}(dW^t\wedge \mathsf{h_{n}} dW\mathsf{\omega_2})$   \\
 \hline
$\mathsf{sp_{2n}}$ &  $W\in \mathrm{Hom}(\CC, \CC^{2n})$ & $dW^t\wedge \mathsf{\omega_{2n}} dW$  \\
 \hline
\end{tabular}
\label{SymplFormTable}
\end{center}
\end{table}

\vspace{-0.5cm}
Notice that in the latter two cases the space of matter fields is `self-dual' due to the existence of invertible tensors on the vector spaces in the nodes: as a result, one can identify each vector space with its dual. To obtain closures of the nilpotent orbits from the data in the table one takes the symplectic quotient w.r.t. the groups in the circular nodes of the quivers. Rectangular nodes are left intact, and the groups displayed therein are global symmetry groups of the orbits.

\section{2D field theories from nilpotent orbits} So far in the preceding two sections we concentrated on the geometry of nilpotent orbits. However, ultimately we are interested in 2D models, which are related to nilpotent orbits via an algorithm of the type described in section~\ref{PCMsect}. In the present section, instead of deriving the e.o.m. of the models from the PCM equation~(\ref{PCMeq}), we will take a different route and describe a way of writing down Lagrangians, from which the corresponding e.o.m. follow.

In the case of minimal nilpotent orbits these Lagrangians may be constructed from the data in the table. First,  consider the $\mathsf{sl_n}$-case, which is a higher-$n$ generalization of the example~\ref{sl2example}. The question is how to pull-back the elementary one-form $\theta_0=\sum_{i=1}^n \,V_i\,dU_i$ to the worldsheet $\Sigma$ to obtain the kinetic term of the model. To this end, one should regard $U_i$ as sections of some line bundle\footnote{On $\Sigma=\CP^1$ non-trivial line bundles $L$ correspond to `instanton' solutions of the sigma model, the degree of $L$ being related to the instanton number.} $L$ over $\Sigma$, and $V_i$ as sections of the dual bundle $L^{-1}\otimes K_{\Sigma}$. In this case the integral
\bea\label{holact}
\mathcal{S}_{\mathrm{hol}}=\int_{\Sigma}\!i\,dz\!\wedge\! d\thickbar{z}\,\sum_{i=1}^n \,V_i\,\thickbar{D}_AU_i\,,
\eea
where $D_A$ is a connection in $L$, is well-defined. Note that the $(U, V)$-variables are complex generalizations of the $(x, p)$-variables used in section~\ref{spheresect} in the case of $T^\ast S^{n-1}$. Variation of the action~(\ref{holact}) w.r.t. the gauge field produces the constraint $\sum_{i=1}^n \,V_i\,U_i=0$, so that the moment map $\upmu=U\otimes V\in \mathsf{sl_n}\otimes K_{\Sigma}$. Thus, the interaction term $\mathrm{Tr}(\upmu\bar{\upmu})$ is a form of type $(1, 1)$ and may be integrated over $\Sigma$. To make contact with the  Gross-Neveu model, it is useful to package the $(U, V)$-variables in a single Dirac spinor
\bea\label{spinor}
\Psi=\begin{pmatrix}U \\ \thickbar{V}\end{pmatrix}\,.
\eea
The total action may then be written as ($\sigma_i$ are the Pauli matrices)
\bea\label{gaugedGN}
\mathcal{S}=\int_{\Sigma}\!i\,dz\!\wedge\! d\thickbar{z}\,\left[\thickbar{\Psi}\slashed{D}_A\Psi+\left(\thickbar{\Psi} {1+\sigma_3\over 2} \Psi\right)\left(\thickbar{\Psi} {1-\sigma_3\over 2} \Psi\right)\right]\,.
\eea
The above action~(\ref{gaugedGN}) is the gauged chiral Gross-Neveu model, albeit for bosonic rather than fermionic fields. One easily checks that the interaction term is proportional to $\mathrm{Tr}(\upmu\bar{\upmu})$, which is the 2D generalization of the Hamiltonian~(\ref{Hmu2}). The resulting e.o.m. are the ones written out in~(\ref{UVeq}) above, so that the system may be shown to be equivalent to the $\CP^{n-1}$ sigma model upon elimination of the $(V, \thickbar{V})$ variables, cf.~(\ref{CPNeq}).

As mentioned earlier, not every nilpotent orbit closure admits a symplectic resolution. We therefore expect that such orbits do not correspond to sigma models\footnote{The orbits that do lead to sigma models with Grassmannian target spaces (in the $\mathsf{u}$-, $\mathsf{o}$- and $\mathsf{sp}$-cases) are briefly discussed in~\cite{BykovKrivorol}.}. For example, let us return to the minimal $\mathsf{sp}_{2n}$-orbit, introduced in section~\ref{PCMsect}. As one can see from the quiver, it is simply $\CC^{2n}/ \mathbb{Z}_2$ (since $\mathsf{O}(1)\simeq \mathbb{Z}_2$). The moment map, which can be computed from the symplectic form in the table, is defined by $\upmu=W\otimes W^t \cdot \omega_{2n}$ and  satisfies $\upmu^2=0$. Choosing $\omega_{2n}=i\sigma_2\otimes \mathds{1}_n$, we may split $W=\begin{pmatrix}U \\ V^t\end{pmatrix}$. In terms of the Dirac spinor~(\ref{spinor}) the action looks as follows:
\bea\label{sp2nGN}
\mathcal{S}=\int_{\Sigma}\!i\,dz\!\wedge\! d\thickbar{z}\,\left[\thickbar{\Psi}\slashed{D}\Psi+\left(\thickbar{\Psi} \Psi\right)^2\right]\,.
\eea
Here again we have introduced the interaction $\mathrm{Tr}(\upmu\thickbar{\upmu})=(\thickbar{W}W)^2=(\thickbar{\Psi} \Psi)^2$, and for this to make sense we should assume that both $U$ and $V$ are sections of $K_{\Sigma}^{1/2}$. Thus, we have arrived at the non-chiral Gross-Neveu model. The action features quartic interactions of the type $|U|^4, |V|^4$, so that neither of the variables may be easily integrated out. This is the practical reason why the relation to sigma models fails in this case.

There is an exception to the above claim for $n=1$. In this case the closure of the nilpotent orbit is $\CC^2/ \mathbb{Z}_2$ and is well-known to admit a resolution\footnote{The hyper-K\"ahler metric on it is the Eguchi-Hanson metric, cf.~\cite{EGH, Perelomov}.} isomorphic to $T^\ast \CP^1$. This is because $\mathsf{Sp(2)}\simeq \mathsf{SL(2)}$, so that we may use an alternative formulation of the same system in terms of the chiral gauged GN-model~(\ref{gaugedGN}) with $n=2$. To prove equivalence, we may resolve the constraint $\sum_{i=1}^2 \,V_iU_i=0$ as $V_i=\lambda \epsilon_{ij} U_j$ with $\lambda \in \CC$ (assuming $U\neq 0$). Substituting this in~(\ref{gaugedGN}), one arrives at the ungauged model~(\ref{sp2nGN}) with $n=1$.

\section{Conclusion}

In the present paper we have described the geometric framework, which conjecturally underpins the relation between 2D sigma models and Gross-Neveu models. As we emphasized, the pivotal role is played here by closures of nilpotent orbits and their resolutions, which naturally lead to cotangent bundles of flag manifolds $G_{\CC}/P$, the latter being target spaces of our sigma models. This is a generalization of the observations in~\cite{Bykov1, Bykov2}, which applies to the case of arbitrary classical Lie group (the exceptional case is of interest is well, but has not been elaborated yet). Besides, in the present paper we also mentioned orbits, whose closures do not admit symplectic resolutions, yet in these cases one also obtains interesting examples of non-chiral Gross-Neveu models (which are, however, unrelated to sigma models).

The nilpotent orbits in question possess a distinguished set of quasi-linear Darboux variables, which may be traced back to a quiver formulation of these orbits. These variables have proven their value in quantum mechanical applications (cf.~\cite{BykovSmilga}), and the intriguing question is whether they are equally useful in the 2D quantum field-theoretic setup. 

\vspace{0.5cm}\noindent
\textbf{Acknowledgments.}

\vspace{0.2cm}
\noindent This paper is dedicated to the memory of my scientific adviser A.A.~Slavnov. I will remember him as a profound yet cheerful person, of immense scientific integrity and dedication to science, and I will always be grateful for his benevolence and support.

\vspace{0.2cm}
\noindent I would like to thank E.~Ivanov, A.~Nersessian, A.~Roslyi, A.~Smilga, P.~Zinn-Justin and members of the I.~R.~Shafarevich seminar, where part of this work was presented, for discussions, useful remarks and suggestions, and especially V.~Krivorol for a thorough reading of the manuscript. This work has been supported by Russian Science Foundation grant \href{https://rscf.ru/en/project/22-72-10122/}{RSCF-22-72-10122}.

\end{document}